\documentclass[12pt]{amsart}
\usepackage{amsmath}
\usepackage{amssymb}
\usepackage{graphicx}
\usepackage[latin1]{inputenc}
\usepackage[T1]{fontenc}
\newtheorem{lemma}{Lemma}

\newtheorem{proposition}{Proposition}
\newtheorem{definition}{Definition}
\newtheorem{remark}{Remark}

\newcommand{\tr}{{\rm Tr }}

\newcommand{\bra}{\langle}
\newcommand{\ket}{\rangle}
\newcommand{\vp}{\varphi}

\newcommand{\N}{\mathbb{N}}

\newcommand{\R}{\mathbb{R}}

\newcommand{\be}{\begin{equation}}
\newcommand{\eeq}{\end{equation}}
\newcommand{\bet}{\begin{equation*}}
\newcommand{\eeqt}{\end{equation*}}
\newcommand{\bea}{\begin{eqnarray}}
\newcommand{\eeqa}{\end{eqnarray}}
\newcommand{\beat}{\begin{eqnarray*}}
\newcommand{\eeqat}{\end{eqnarray*}}
\newcommand{\goesto}{\rightarrow}
\newcommand{\goto}{\goesto}
\newcommand{\h}[1]{\mathcal{#1}}
\newcommand{\hil}{\mathcal{H}}

\newcommand{\cc}[1]{\overline{#1}}

\begin{document}
\title[Phase space observables and eight-port homodyne detector]{A note on the measurement of phase space observables with an eight-port homodyne detector}
\author{J. Kiukas}
\address{Jukka Kiukas,
Department of Physics, University of Turku,
FIN-20014 Turku, Finland}
\email{jukka.kiukas@utu.fi}
\author{P. Lahti}
\address{Pekka Lahti,
Department of Physics, University of Turku,
FIN-20014 Turku, Finland}
\email{pekka.lahti@utu.fi}

\begin{abstract}
It is well known that the Husimi Q-function of the signal field can actually be measured by the eight-port homodyne detection technique,
provided that the reference beam (used for homodyne detection) is a very strong coherent field so that it can be treated classically, see e.g. 
\cite{Leonhardt}. Using recent rigorous results  on the quantum theory of homodyne detection observables \cite{homodyne}, we show
that any phase space observable, and not only the Q-function, can be obtained as a high amplitude limit of the
signal observable actually measured by an eight-port homodyne detector. The proof of this fact does not involve any classicality assumption.  
\end{abstract}

\maketitle

\section{Introduction}

Covariant phase space observables, as positive operator measures, play an important role in the foundations of quantum mechanics. 
In particular, their importance  for the approximate joint measurements of position and momentum observables has long been recognized\footnote{See,
for instance,  the monographs \cite{Davies,Holevo,OQP,Stulpe}.}  and it has  recently been shown \cite{Werner04} that for any approximate joint measurement of position and
momentum of a quantum object there is a covariant phase space observable with improved degrees of approximations.\footnote{For details of these 
concepts as well as for a further analysis of these 
results, see the above quoted work of Werner \cite{Werner04} as well as the subsequent developments
\cite{CHT05,BHL07,BP07}.} 
The  mathematical structure of  the covariant phase space observables is also completely known: they correspond one-to-one onto the positive operators of 
trace one (acting on the Hilbert space of the quantum object in question), and they have an operator density defined by the Weyl operators 
and the positive trace-one operator in question, see eq. (\ref{phasespaceobs}) below.\footnote{This result is due to Holevo \cite{Holevo79} and Werner \cite{Werner84},
alternative proofs with different techniques were recently given in \cite{Cassinelli03} and \cite{Kiukas06}.} 
Moreover, the realization of the covariant phase space observables (associated with one-dimensional projection operators) as sequential position-momentum
measurements, with the first measurement as an approximately repeatable measurement, has recently been demonstrated in \cite{CHT07} following the
pioneering work of Davies \cite{Davies70}. What remains then is the question of the experimental implementation of the phase space observables.

It is well-known that using the classical approximation of high amplitude reference beam,
the Husimi Q-function (i.e. the phase space observable generated by the vacuum state operator)
is obtained as the measured observable for the so called eight-port homodyne detector (see \cite{Leonhardt}, and
\cite[p. 147-155]{LeonhardtII}).
In this paper we show, using the recent results \cite{homodyne} on the balanced homodyne detection observables,
that any covariant phase space observable can be obtained, in a mathematically rigorous sense, as the high
amplitude limit of signal observables determined by the eight-port homodyne detection scheme.

\section{The eight-port homodyne detector}

\begin{figure}
\includegraphics[width=15cm]{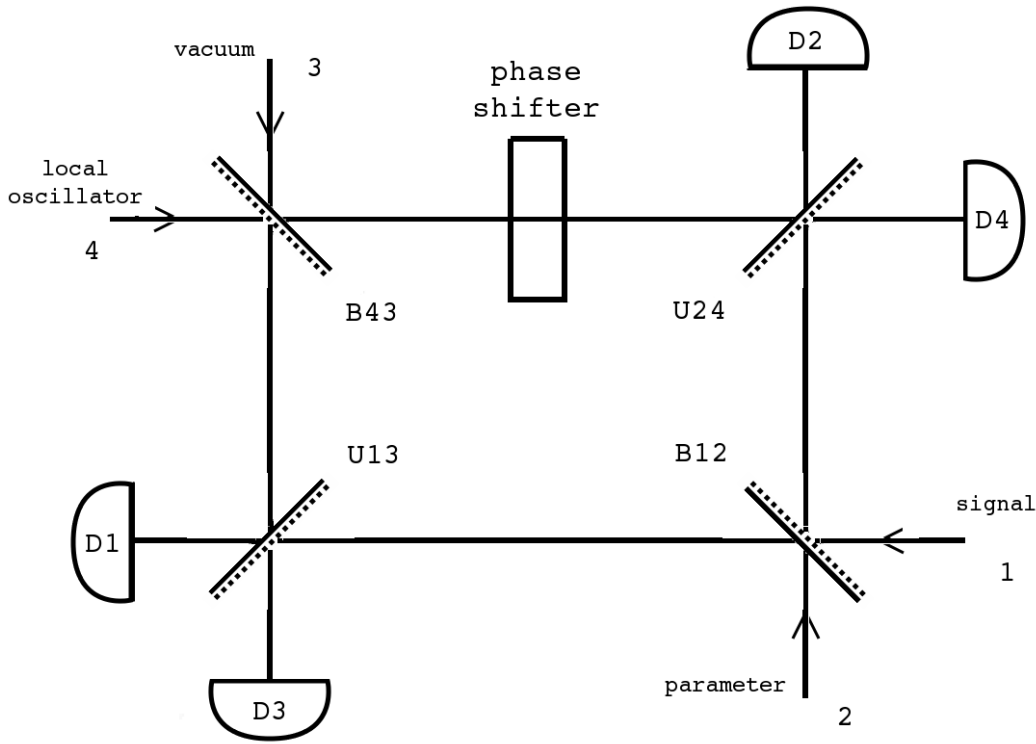}
\caption{The eight-port homodyne detector}
\end{figure}

%

The detector involves four modes as indicated in the picture, and we will denote the associated
(complex separable) Hilbert spaces accordingly by $\hil_1$, $\hil_2$, $\hil_3$, $\hil_4$. Mode 1 corresponds to
the signal field (i.e. the object system with respect to which the measured observable will be interpreted),
the input state for mode 2 serves as a parameter which determines (as will be seen below)
the phase space observable to be measured,
and mode 4 is the reference beam in a coherent state. (The input for mode 3 is left empty, corresponding to
the vacuum state.)

We fix the photon number bases $\{|n\ket\mid n\in \N\}$ for each $\hil_i$, so that the annihilation operators
$a_j$, as well as the quadratures
$Q_j = \tfrac {1}{\sqrt 2} (\cc{a_j^*+a_j})$, $P_j = \tfrac {1}{\sqrt 2} i(\cc{a_j^*-a_j})$ and photon number
operators $N_j=a_j^*a_j$ are defined for each mode $j=1,2,3,4$. The bar above denotes the closure of an operator.
We will also sometimes use the coordinate
representation for each $\hil_j$ (i.e $\hil_j \simeq L^2(\R)$, with $|n\ket$ associated with $n$th Hermite function).
Then $Q_j$ and $P_j$ act as usual position and momentum operators: $(Q_j\psi)(x) = x\psi(x)$,
and $(P_j\psi)(x) =-i\frac {d\psi}{dx}(x)$, $\psi\in L^2(\R)$.
For any selfadjoint operator $A$ on a Hilbert space, we let $P^A$ denote its spectral measure.

The photon detectors $D_j$ shown in the picture are considered to be ideal, so that each detector $D_j$
measures the sharp photon number $N_j$. The phase shifter in mode 4 is represented by the unitary operator
$e^{i\phi N_4}$, where $\phi$ is the shift.

There are four 50-50-beam splitters $B_{12}$, $B_{43}$, $U_{13}$, $U_{24}$,
each of which is defined by its acting in
the coordinate representation (see e.g. \cite{Leonhardt}):
$$
L^2(\R^2)\ni \Psi\mapsto \big(\,(x_1,x_2)\mapsto \Psi(\tfrac 1 {\sqrt 2} (x_1+x_2),\tfrac 1 {\sqrt 2} (-x_1+x_2))\,\big)\in L^2(\R^2),
$$
Under this transform, the coherent states change according to
\begin{equation}\label{coherent}
|\alpha\ket\otimes |\beta\ket \mapsto |\tfrac 1 {\sqrt 2} (\alpha-\beta)\ket\otimes |\tfrac 1 {\sqrt 2} (\alpha+\beta)\ket.
\eeq
In the picture, the dashed line in each beam splitter indicates the input port of the ''primary mode'',
i.e. the mode associated with the
first component of the tensor product $L^2(\R)\otimes L^2(\R)\simeq L^2(\R^2)$ in the above description.
The beam splitters $B_{12}$, $B_{43}$, $U_{13}$ and $U_{24}$ are indexed so that the first index
indicates the primary mode.
(For instance, $B_{43}$ acts such that an input two-mode state $\vp\otimes\psi\in \hil_3\otimes\hil_4$
is interpreted as $\Psi(x_1,x_2):=\psi(x_1)\vp(x_2)$ in the above definition.)

The unitary operator describing the entire transform caused by the combination of the beam splitters together with the phase shifter is
$$
(U_{13}\otimes U_{24})e^{i\phi I_{123}\otimes N_4}(B_{12}\otimes B_{43}).
$$
Here it is understood that
$$
\hil_1\otimes \hil_2\otimes\hil_3\otimes \hil_4\simeq (\hil_1\otimes \hil_2)\otimes(\hil_3\otimes \hil_4) \simeq (\hil_1\otimes\hil_3)\otimes (\hil_2\otimes \hil_4),
$$
and we will freely use these isometries, without explicit indication, when the ordering is clear from the context.

Let $|\sqrt 2 z\ket$ be the coherent input state for mode 4. We choose to detect the scaled number
differences $\tfrac 1 {|z|} N_{13}^-$ and $\tfrac 1 {|z|} N_{24}^-$, where
$$
N_{13}^-:=\cc {I_1\otimes N_3-N_1\otimes I_3}, \ \ N_{24}^-:=\cc {I_2\otimes N_4-N_2\otimes I_4},
$$
so that the detection statistics are described by the biobservable
$$
(X,Y)\mapsto P^{|z|^{-1} N_{13}^-}(X)\otimes P^{|z|^{-1} N_{24}^-}(Y)
$$
acting on the entire four-mode field.

Let the input states of modes 1 and 2 be $T$ and $S$, respectively, so that the input for the four-mode field is
$T\otimes S\otimes | 0\ket\bra 0|\otimes |\sqrt{2} z\ket\bra\sqrt{2} z|$ (with the natural
ordering of tensor products). The action of $B_{43}$, as well as the phase shifter can be calculated explicitly
in terms of coherent states:
$$
(I_3\otimes e^{i\phi N_4})B_{43}(|0\ket\otimes |\sqrt 2 z\ket) = |z\ket\otimes |e^{i\phi}z\ket.
$$
(Notice that here mode 4 is the primary mode, so that equation \eqref{coherent} is applied
with the tensor product order reversed.) Now the state of the system after the transform can be written as
$$ 
U_{13}\otimes U_{24} W_{T,S,z,\phi} U_{13}^*\otimes U_{24}^*,
$$
where
$$
W_{T,S,z,\phi}:=B_{12}(T\otimes S)B_{12}^*\otimes |z\ket\bra z|\otimes |ze^{i\phi}\ket\bra ze^{i\phi}|.
$$
(Here the tensor product is written in the original order.)

The detection statistics are given by the probability bimeasures
$$
p_T^{S,z,\phi}(X,Y) := \tr [U_{13}\otimes U_{24} W_{T,S,z,\phi} U_{13}^*\otimes U_{24}^* P^{|z|^{-1} N_{13}^-}(X)\otimes P^{|z|^{-1} N_{24}^-}(Y)].
$$
Since $\frac 1 {\sqrt 2} U_{13}^*N_{13}^-U_{13} = A_{13}$, and $\frac 1 {\sqrt 2} U_{24}^*N_{24}^-U_{24} = A_{24}$,
where e.g. $A_{13} = \frac 1 {\sqrt 2} \cc {(a_1\otimes a_3^*+a_1^*\otimes a_3)}$, we have simply
$$
p_T^{S,z,\phi}(X,Y) = \tr [W_{T,S,z,\phi} P^{\sqrt 2|z|^{-1}A_{13}}(X)\otimes P^{\sqrt 2|z|^{-1}A_{24}}(Y)].
$$ 
Let $P^z:\h B(\R^2)\to L(\hil_1\otimes \hil_2\otimes \hil_3\otimes \hil_4)$ denote the unique spectral measure
extending the biobservable
$(X,Y)\mapsto P^{|z|^{-1}A_{13}}(X)\otimes P^{|z|^{-1}A_{24}}(Y)$, so that
$$
p_T^{S,z,\phi}(X,Y) = \tr [W_{T,S,z,\phi}P^z(\tfrac 1{\sqrt 2} (X\times Y))]
$$
for any $X,Y\in \h B(\R)$.

In our measurement, the input state $S$ of mode 2, as well as the coherent input state $|\sqrt 2 z\ket$ of mode 4,
and the phase shift $\phi$,
are regarded as fixed parameters, while $T$ is the state of the object system, i.e. mode 1. Accordingly, we define
an observable $G^{z,S,\phi}:\h B(\R^2)\to L(\hil_1)$ via
\be\label{theobservable}
\tr [TG^{z,S,\phi}(Z)] = \tr [W_{T,S,z,\phi}P^z(\tfrac 1{\sqrt 2}Z)], \ Z\in\h B(\R^2).
\eeq
(Since $\tr [W_{T,S,z,\phi}P^z(\tfrac 1{\sqrt 2}Z)]= \tr [T\otimes S\otimes |z\ket\bra z|\otimes |ze^{i\phi}\ket\bra ze^{i\phi}| B_{12}^*\otimes I_{34}P^z(\tfrac 1{\sqrt 2}Z)B_{12}\otimes I_{34}]$, it follows by standard duality and convergence arguments that the
observable $G^{z,S,\phi}$ exists and is uniquely determined by the above formula.)
{\bf The observable measured by the eight port homodyne detector is thus $G^{z,S,\phi}$.}

In order to consider the limit $|z|\goesto \infty$, we need to express $G^{z,S,\phi}$ in terms of the single homodyne
detector observables $E_{1}^z$ and $E_{2}^{ze^{i\phi}}$, where 
$E_{1}^z(X):= (V^{13}_z)^*P^{|z|^{-1}A_{13}}(X)V_z^{13}$, $V^{13}_z:\hil_1\to \hil_1\otimes \hil_3$ is the isometry
$\phi\mapsto \phi\otimes |z\ket$, and $E_{2}^{ze^{i\theta}}(Y)$ is defined in an analogous way by using
$A_{24}$ and $V_{ze^{i\phi}}^{24}$ (see \cite{homodyne}).  We let $\h S(\hil_i)$ denote the set of states (that is, positive operators of trace one) of the mode
described by the Hilbert space $\hil_i$, $i=1,2,3,4$. 

\begin{lemma}\label{Glemma}
For any $T\in \h S(\hil_1)$, $X,Y\in \h B(\R)$, and an $S\in \h S(\hil_2)$,
we have
\be\label{traceform}
\tr[TG^{z,S,\phi}(X\times Y)] = \tr [B_{12}(T\otimes S)B_{12}^* E_1^z(\tfrac 1{\sqrt 2}X)\otimes E_2^{e^{i\phi}z}(\tfrac 1{\sqrt 2}Y)].
\eeq
\end{lemma}
\begin{proof} Let $X,Y\in \h B(\R)$.
Assume first that $T=P[\vp]$ and $S=P[\psi]$ for some unit vectors $\vp\in \hil_1$, $\psi\in\hil_2$. Write
$$B_{12}(\vp\otimes \psi) = \sum_{n,m=0}^\infty c_{nm} \eta^1_n\otimes\eta_n^2,$$
where $\{\eta_n^i\mid n \in \N\}$ is an orthonormal basis for $\hil_i$, $i=1,2$. This series converges in
the norm of $\hil_1\otimes \hil_2$, so by using the definition \eqref{theobservable}, we get
\beat
&& \tr[TG^{z,S,\phi}(\sqrt 2 (X\times Y))] = \Big{\langle} B_{12}(\vp\otimes\psi)\otimes |z\ket\otimes |e^{i\phi}z\ket\Big| P^z(X\times Y)
B_{12}(\vp\otimes\psi)\otimes |z\ket\otimes |e^{i\phi}z\ket\,\Big\rangle\\
&=& \sum_{n,m,n',m'=0}^\infty \cc{c_{nm}}c_{n'm'}
\Big{\langle} \eta_n^1\otimes\eta_m^2\otimes |z\ket\otimes |ze^{i\phi}\ket \Big|P^{|z|^{-1}A_{13}}(X)\otimes P^{|z|^{-1}A_{24}}(Y)
\eta_{n'}^1\otimes\eta_{m'}^2\otimes |z\ket\otimes |ze^{i\phi}\ket\, \Big\rangle\\
&=& \sum_{n,m,n',m'=0}^\infty \cc{c_{nm}}c_{n'm'} \bra \eta_n^1 |(V_z^{13})^*P^{|z|^{-1}A_{13}}(X)V_z^{13}\,\eta_{n'}^1\ket
\bra \eta_m^2 |(V_{ze^{i\phi}}^{24})^*P^{|ze^{i\phi}|^{-1}A_{24}}(Y)V_{ze^{i\phi}}^{24}\,\eta_{m'}^2\ket\\
&=& \sum_{n,m,n',m'=0}^\infty \cc{c_{nm}}c_{n'm'} \bra \eta_n^1\otimes \eta_m^2|E_1^z(X)\otimes E_2^{ze^{i\phi}}(Y)\,\eta_{n'}^1\otimes \eta_{m'}^2\ket\\
&=& \bra B_{12}(\vp\otimes\psi) |E_1^z(X)\otimes E_2^{ze^{i\phi}}(Y) B_{12}(\vp\otimes\psi)\ket
=\tr [B_{12}(T\otimes S)B_{12}^* E_1^z(X)\otimes E_2^{e^{i\theta z}}(Y)].
\eeqat
As for the general case, note that both sides of \eqref{traceform} are trace norm continuous when regarded
as functions of $T$ (for fixed $S$), and the same is true when they are regarded as functions of $S$
for fixed $T$. Hence, by using the spectral resolutions for $S$ and $T$ (which are trace norm convergent weighted
sums of one-dimensional projectors), one establishes the claim.
\end{proof}

\section{The high amplitude limit}

Now we proceed to describe the limit $|z|\goesto \infty$, in the case where $z=r>0$, and $\phi=\frac {\pi}{2}$.
In order to simplify the notation, we will drop the subscript for mode 1 operators (e.g. $Q:=Q_1$) from now on.
We need the following general definition, which we introduced in \cite{homodyne}.

\begin{definition}\label{weakdef}\rm Let $\hil$ be a Hilbert space, $\Omega$ a metric space with Borel $\sigma$-algebra $\h B(\Omega)$, and $E^n:\h B(\Omega)\to L(\hil)$ a semispectral measure for each $n\in \N$. We say that the sequence
$(E^n)$ converges to a semispectral measure $E:\h B(\Omega)\to L(\hil)$ \emph{weakly in the sense of probabilities}, if
$$ \lim_{n\goto\infty} E^n(X) = E(X)$$ in the weak operator topology, for all $X\in \h B(\Omega)$ such that $E(\partial X)=0$, where $\partial X$ is the boundary of the set $X$. (Recall that $\partial X$ is the intersection of the
closures of $X$ and its complement.)
\end{definition}

We used this definition in the context where $\Omega = \R$, and proved that for any $\theta\in [0,2\pi)$, and
any sequence $(r_n)$ of positive numbers converging to infinity,
the sequence $(E^{r_ne^{i\phi}})_{n\in\N}$ of homodyne detector observables converges to $P^{Q_\theta}$
weakly in the sense of probabilities, where $Q_\theta=e^{i\theta N} Qe^{-i\theta N}$ is the rotated quadrature \cite[p. 17]{homodyne}. Since the spectral measure of each quadrature is absolutely continuous with the Lebesgue measure $\lambda$, the condition $P^{Q_\theta}(\partial X)=0$ for an $X\in \h B(\R)$ is reduced to the condition $\lambda(\partial X)=0$.
Now we need only the particular choices $\phi = 0$ and $\phi=\frac {\pi}{2}$,
when the limiting observables are the spectral measures of $Q$ and $P$, respectively.

For any positive operator $S\in L(\hil_1)$ of trace one, let $E^S:\h B(\R^2)\to L(\hil_1)$ denote the phase space observable
generated by $S$, i.e.
\begin{equation}\label{phasespaceobs}
E^S(Z) = \frac {1}{2\pi} \int_Z W(q,p)SW(q,p)^* \, dqdp,
\end{equation}
where $W(q,p)= e^{i\frac 12 qp} e^{-iqP}e^{ipQ}$, $q,p\in \R$ are the Weyl operators.
Let $C:\hil_2\to\hil_1$ denote the (antiunitary) conjugation operator $\psi\mapsto (x\mapsto \cc{\psi(x)}))$
Notice that here $C$ is interpreted a map from $\hil_2\simeq L^2(\R)$ to $\hil_1\simeq L^2(\R)$. Since $C$ is antiunitary,
it follows that for any positive operator $S\in L(\hil_2)$ of trace one, the map $CSC^{-1}$ is a bounded linear positive
operator in $\hil_1$ with unit trace.

The following well-known result can be found e.g from \cite[p. 195-196]{OQP}. We reproduce it here so as to
make sure that the notations fit together correctly.

\begin{lemma} \label{faasilemma} Let $S\in L(\hil_2)$ be a positive operator of trace one, and
$T\in \h S(\hil_1)$. Then
\be\label{phaserelation}
\tr [TE^{CSC^{-1}}(X\times Y)] = \tr [B_{12}(T\otimes S)B_{12}^* P^Q(\tfrac 1{\sqrt 2}X)\otimes P^{P_2}(\tfrac 1{\sqrt 2}Y)],
\ \ X,Y\in \h B(\R).
\eeq
\end{lemma}
\begin{proof} We have $P_2= F^{-1}Q_2F$, where $F$ is the Fourier-Plancherel operator. If $\vp\in L^2(\R)\cap L^1(\R)$,
the operator $F$ acts as the Fourier transform:
$$
[F\vp](y) = \frac {1}{\sqrt{2\pi}} \int e^{-iyx} \vp(x)\, dx.
$$
The relation
$$
\frac {1}{\sqrt {2\pi}} \bra W(q,p)C\eta|\xi\ket
= e^{\frac 12i qp}[F[\eta(\cdot -q)\xi]](p)
$$
holds for all $\xi\in \hil_1$, $\eta\in \hil_2$, $\|\eta\|=1$, all $p\in \R$ and almost all $q\in\R$, with the function
$x\mapsto \eta(x-q)\xi(x)$ belonging to $L^1(\R)\cap L^2(\R)$ for almost $q\in \R$
(see e.g. \cite[p. 47, 49]{Stulpe}).

First, let $S=P[\psi]$ for some unit vector $\psi\in \hil_2$, and let $\vp\in \hil_1$.
Because of the above relation, we have
\beat
\sqrt 2 \frac {1}{\sqrt {2\pi}} \bra W(\sqrt 2 q,\sqrt 2 p)C\psi|\vp\ket
&=& \sqrt 2 e^{i qp}[F[\psi(\cdot -\sqrt 2 q)\vp](\sqrt 2 p)\\
&=& \sqrt 2 \frac {1}{\sqrt {2 \pi}}\int e^{-ip(\sqrt 2 x-q)} \vp(x)\psi(x-\sqrt 2 q) \, dx\\
&=& \frac {1}{\sqrt {2 \pi}}\int e^{-ipy} \vp(\tfrac {1}{\sqrt 2} (q+y))\psi(\tfrac {1}{\sqrt 2} (-q+y)) \, dy\\
&=& F[[B_{12}(\vp\otimes \psi)](q,\cdot)](p)
\eeqat
for all $p\in \R$ and almost all $q\in \R$, and, subsequently
\begin{align*}
\tr [B_{12}(P[\vp]\otimes P[\psi]) & B_{12}^* P^Q(\tfrac 1{\sqrt 2}X)\otimes P^{P_2}(\tfrac 1{\sqrt 2}Y)] \\
&= \bra (I\otimes F) B_{12}(\vp\otimes \psi)| P^Q(\tfrac 1{\sqrt 2}X)\otimes P^{Q_2}(\tfrac 1{\sqrt 2}Y)(I\otimes F) B_{12}(\vp\otimes \psi)\ket\\
&= \int_{\tfrac 1{\sqrt 2}(X\times Y)} |[(I\otimes F) B_{12}(\vp\otimes \psi)](q,p)|^2 \, dqdp\\
&= \int_{\tfrac 1{\sqrt 2}(X\times Y)} |[F[B_{12}(\vp\otimes\psi)](q,\cdot)](p)|^2\, dqdp\\
&= \frac 1 {2\pi}\int_{\tfrac 1{\sqrt 2}(X\times Y)} 2 |\bra W(\sqrt 2 q,\sqrt 2 p) C\psi|\vp\ket|^2\, dqdp\\
&= \frac 1 {2\pi}\int_{X\times Y} |\bra W(q,p) C\psi|\vp\ket|^2\, dqdp\\
&= \bra \vp |E^{P[C\psi]}(X\times Y)\vp\ket.
\end{align*}

Hence, we have shown that \eqref{phaserelation} holds when $S=P[\psi]$ and $T=P[\vp]$. Since
$\tr[TE^{S'}(Z)] = \tr[S'E^T(-Z)]$ for any $Z\in \h B(\R^2)$ and a positive operator $S'$ of trace one, the linear maps
$T\mapsto \tr[TE^{CSC^{-1}}(Z)]$ and $S\mapsto \tr[TE^{CSC^{-1}}(Z)]$ are trace-norm continuous. Clearly
also the right hand side of \eqref{phaserelation} depends linearly and continuously on both $S$ and $T$, so that
the proof is completed by applying the spectral representations for $S$ and $T$.
\end{proof}

The convergence in the high amplitude limit is characterized by the following Proposition, where Definition \ref{weakdef}
is used in the case where $\Omega = \R^2$.

\begin{proposition} Let $S\in L(\hil_2)$ be any positive operator of trace one, and let
$(r_n)$ be any sequence of positive numbers converging to infinity. Then
the sequence $(G^{r_n,S,\tfrac{\pi}{2}})_{n\in \N}$ converges to the phase space observable
$E^{CSC^{-1}}$ weakly in the sense of probabilities.
\end{proposition}
\begin{proof}
Fix $(r_n)$ to be a sequence of positive numbers converging to infinity.
Let $X,Y\in \h B(\R)$ be such that $\lambda(\partial X)=\lambda(\partial Y)=0$, where $\lambda$ is the Lebesgue measure
of $\R$. According to the discussion at the beginning of the section,
$(E_1^{r_n}(\tfrac 1{\sqrt 2}X))_{n\in \N}$ converges to
$P^Q(\tfrac 1{\sqrt 2}X)$ in the weak operator topology of $L(\hil_1)$, and
$(E_2^{r_n i}(\tfrac 1{\sqrt 2}Y))_{n\in \N}$ converges to $P^{P_2}(\tfrac 1{\sqrt 2}Y)$
in the weak operator topology of $L(\hil_2)$. Since the norms of all these operators are bounded by $1$, it follows that
the tensor product operator sequence $(E_1^{r_n}(\tfrac 1{\sqrt 2}X)\otimes E_2^{r_n i}(\tfrac 1{\sqrt 2}Y))_{n\in \N}$
converges to $P^Q(\tfrac 1{\sqrt 2}X)\otimes P^{P_2}(\tfrac 1{\sqrt 2}Y)$ in the weak operator topology of $\hil_1\otimes \hil_2$. The boundedness
of the operator norms further implies that the latter sequence actually converges ultraweakly.
Hence, it follows from Lemma \ref{Glemma} and \ref{faasilemma} that for any $T\in \h S(\hil_1)$, we have
$$
\lim_{n\goesto\infty} \tr [TG^{r_n,S,\tfrac{\pi}{2}}(X\times Y)] =
\tr [B_{12}(T\otimes S)B_{12}^* P^Q(\tfrac 1{\sqrt 2}X)\otimes P^{P_2}(\tfrac 1{\sqrt 2}Y)]
= \tr [TE^{CSC^{-1}}(X\times Y)].
$$
Since the family
$$\{ X\times Y \mid X,Y\in \h B(\R), \, \lambda(\partial X)=\lambda(\partial Y) = 0\}$$
is closed under finite intersections (note that $\partial(X_1\cap X_2)\subset \partial(X_1)\cup\partial(X_2)$,
$X_1,X_2\in \h B(\R)$), and
includes a neighborhood base (for the usual topology of $\R^2$) of any point $x\in \R^2$, it follows from
\cite[Corollary 1, p. 14]{Billingsley} that for any $T\in \h S(\hil_1)$, the sequence
$(Z\mapsto \tr [TG^{r_n,S,\tfrac{\pi}{2}}(Z)])_{n\in \N}$ of probability measures on $\R^2$ converges weakly to
the probability measure $Z\mapsto \tr[TE^{CSC^{-1}}(Z)]$. (Recall that a sequence $(\mu_n)$ of probability measures
on $\R^2$ is said to converge weakly to a probability measure $\mu$, if $\lim_n\int f \,d\mu_n = \int f \, d\mu$
for each bounded continuous real function $f$ \cite[p.11]{Billingsley}.)
An application of Proposition 10 of \cite{homodyne} now completes the proof.
\end{proof}
\begin{remark}\rm
Since any phase space observable $E^S$ is absolutely continuous with respect to the Lebesgue measure, it follows that
$$w-\lim_{n\goesto \infty} G^{r_n,S,\tfrac{\pi}{2}}(Z) = E^{CSC^{-1}}(Z)$$
for any $Z\in \h B(\R^2)$ with $\lambda^2(\partial Z)= 0$, where $\lambda^2$ is the two-dimensional Lebesgue measure.
\end{remark}

\end{document}